\documentclass[prd,nofootinbib,preprintnumbers,twocolumn,showpacs,floatfix,superscriptaddress]{revtex4-1}

\usepackage{amsmath}
\usepackage{amssymb}
\usepackage{graphicx}
\usepackage{color}
\usepackage{hyperref}
\usepackage{dsfont}
\usepackage{array}

\newtheorem{lemma}{Lemma}

\newcolumntype{L}{>{\centering\arraybackslash}m{2.5cm}}

\newcommand\GeV{\text{GeV}}

\def\eps{\varepsilon}
\def\unit{\mathds{1}}
\def\O{\mathcal{O}}

\def\vev#1{\left\langle #1\right\rangle}

\newcommand{\be}{\begin{equation}}
\newcommand{\ee}{\end{equation}}
\newcommand{\bd}{\begin{displaymath}}
\newcommand{\ed}{\end{displaymath}}
\newcommand{\bea}{\begin{eqnarray}}
\newcommand{\eea}{\end{eqnarray}}
\newcommand{\nn}{\nonumber}

\def\endproof{\begin{flushright}
$\Box$
\end{flushright}}

\begin{document}

\title{Flavour Structure of GUTs and Uncertainties in Proton Lifetime Estimates
}

\author{Helena Kole\v{s}ov\'a}
\email{helena.kolesova@uis.no}
\affiliation{Faculty of Science and Technology, University of Stavanger,
N-4036 Stavanger, Norway\\[1ex]}
\affiliation{Institute of Particle and Nuclear Physics,
Faculty of Mathematics and Physics,
Charles University in Prague, V Hole\v{s}ovi\v{c}k\'ach 2,
180 00 Praha 8, Czech Republic\\[1ex]}

\author{Michal Malinsk\'y}
\email{malinsky@ipnp.troja.mff.cuni.cz}
\affiliation{Institute of Particle and Nuclear Physics,
Faculty of Mathematics and Physics,
Charles University in Prague, V Hole\v{s}ovi\v{c}k\'ach 2,
180 00 Praha 8, Czech Republic\\[1ex]}


\begin{abstract}
We study the flavour aspects of proton lifetime estimates in simple Grand unified models paying particular attention to their inherent fragility due to the notorious lack of control of some of the key parameters governing the relevant hard-process amplitudes. Among these, the theoretical uncertainties in the flavour structure of the baryon and lepton number violating charged currents due to the potential higher-order effects afflicting the matching of the underlying Yukawa couplings to the low-energy data often play a prominent role.
Focusing on the minimal variants of the most popular unified models we study the potential instabilities of the corresponding proton lifetime estimates based on the renormalizable-level Yukawa fits with respect to the Planck-scale induced flavour effects. In particular, we perform a detailed numerical analysis of all minimal $SO(10)$ Yukawa sector fits available in the literature and show that the proton lifetime estimates based on these inputs exhibit a high degree of robustness with respect to moderate-size perturbations, well within the expected ``improvement window'' of the upcoming proton decay searches. 
\noindent
\end{abstract}
\pacs{{12.60.Fr,14.60.Pq,14.80.Va}}

\maketitle
\section{Introduction\label{sect:intro}}
Mortality of protons is one of the most prominent smoking-gun signals of the idea that strong and electroweak interactions may be just different facets of a unified gauge dynamics at a super-large (YeV) scale. Since its conception in mid 1970's~\cite{GeorgiGlashow} there has been a number of attempts to estimate proton lifetime at vastly different levels of accuracy characterized, namely, by the steadily improving quality of the input data, progress in the field theory calculation techniques, better understanding of the hadronic matrix elements and so on. 

With the upcoming generation of  dedicated experimental searches planned with the Hyper-K and/or DUNE facilities~\cite{Abe:2011ts,Acciarri:2015uup}, which should be able to push the current lower limits (e.g., $\tau_p>1.4\times 10^{34}\,$y in the ``golden'' $p\to \pi^0 e^+$ channel~\cite{SuperK2012}) by as much as one order of magnitude, the importance of a good quality prediction becomes particularly pronounced. 

From this perspective, the current status of the theory affairs is far from satisfactory. Barring the non-perturbative nature of the hadronic layer, even at the level of the underlying ``hard'' processes, i.e.,  with amplitudes featuring quarks rather than hadrons as initial and final states, good quality calculations turn out to be an endeavour of enormous complexity.  Indeed,  the simple structure of the basic baryon and lepton number violating (BLNV) vector currents\footnote{In this study we will focus predominantly on the vector-boson-mediated amplitudes as those mediated by the colored scalars are often sub-leading due to the usual suppression of their couplings to the first-generation quarks and leptons.} even in the minimal  Georgi-Glashow $SU(5)$ model~\cite{GeorgiGlashow}, as simple as it reads,
\be \label{LagBLNV}
{\cal L}_{SU(5)}\propto \overline{u^{c}}\gamma^{\mu}X^{\dagger}_{\mu}Q+\overline{e^{c}}\gamma^{\mu}X_{\mu}Q+\ldots\,,
\ee
encompasses a great deal of arbitrariness (in the mass of the leptoquark $X_{\mu}$, to be identified with the GUT scale $M_{G}$, and, in particular, in the flavour structure emerging when these currents are recast in the quark and lepton mass basis) that may be only partially reflected in the currently accessible low-energy observables. 

Concerning the relative impact of uncertainties in these basic parameters on the proton lifetime estimates, the most critical of these is the value of $M_{G}$ which is determined from the requirement of a proper coalescence of the three Standard Model (SM) gauge couplings at (about) that scale. To this end, note that the logarithmic nature of the gauge running makes even a small error in the low-energy boundary (or high-scale matching) conditions propagate into $M_{G}$ exponentially. This, in turn, calls for\footnote{Let us note that the uncertainty in the low scale value of the strong coupling induces a bigger error than omitting the three and higher loop contributions to the gauge running; hence, at the moment, two-loop precision is the maximum one can do.} higher-loop account of the running effects including the appropriate-level threshold corrections both at $M_{Z}$ as well as at $M_{G}$ (and other intermediate scales, if present); needless to say, this is a highly technically demanding task in practice.

Second in the row is the high degree of uncertainty in the flavour structure of the baryon and lepton number violating (BLNV) charged currents which is namely due to the generic lack of low-energy access to the current-to-mass-basis rotations in the sector of right-handed fermions (as the CKM and PMNS matrices are combinations of the left-handed ones only). If no extra information (such as, e.g., symmetry features of some of the fermionic mass matrices) is available, the total freedom in these unitary transformations is usually enough to spread the outcome of the proton lifetime calculation over many orders of magnitude.

In this respect, it is remarkable that the classical show-stopper of the past, namely, the uncertainties in the hadronic matrix elements, have recently got tamed to such a degree (with typical errors pulled down to few tens of percent) that, nowadays, they can be safely placed as only third in the row, see, e.g.,~\cite{Aoki:2017puj} and references therein.

With this basic hierarchy at hand, one can perform a simple classification of the robustness of the most commonly followed strategies in predicting proton lifetime:  1) The first attempt usually consists in the renormalization group (RG) analysis of the gauge unification constraints which provides information about~$M_{G}$ but often ignores the flavour structure of the BLNV vector currents, typically because the scalar sector of the model is not fully fixed or analyzed. Hence, the uncertainties of thus obtained proton lifetime estimates are generally huge, stretching over many orders of magnitude.  This, however, to a large extent hinders the prospects of discrimination among different scenarios. 2) Sometimes, a great deal of information may be derived from the symmetry features of the effective fermion mass matrices even without performing their detailed fit (usually very demanding), see, e.g.,~\cite{Dorsner:2004xa,Dorsner:2004jj,FileviezPerez:2004hn}. In specific scenarios like, e.g., in the minimal realistic SU(5) models, this may be enough to draw rather accurate conclusions about at least some of the partial decay widths (though often not for the ``golden'' channel $p\to \pi^0 e^+$); the potential to discriminate among such models is obviously much higher then.  3)~The ultimate achievement would be clearly a full-fledged combined analysis of the running together with a detailed Yukawa sector fit. This, however, is very difficult in practice and only very few such attempts have been undertaken in the literature, see, e.g.,~\cite{Babu:2015bna}. 

Nevertheless, even in the most favourable situation of case 3) above there is often an extra source of large and essentially irreducible uncertainties plaguing any proton lifetime estimate obtained in the realm of the simplest renormalizable models, namely, the effects of the higher dimensional effective operators, especially those including the scalar field(s) (to be denoted $S^l$) responsible for the GUT-scale symmetry breaking, i.e., the ones with the vacuum expectation values (VEVs) of the order of $M_{G}$. 

At first glance, there is a number of such structures to be considered
 at the $d=5$ level (with the ordering reflecting their expected ``nuisance'' power), e.g.,
\bea
{\cal O}_{1}&\equiv&\kappa^l_{1} X^{\mu\nu}X_{\mu\nu}S^l/M_{Pl}\nn\\
{\cal O}_{2}&\equiv&\kappa_{2}^{ij,kl} {f_{i}}f_{j}H^k S^l/M_{Pl}\label{OpsList}\\
{\cal O}_{3}&\equiv&\kappa_{3}^{ij,l} {f_{i}}\,\,\slash\!\!\!\!D f_{j}S^l/M_{Pl}\nn\\
{\cal O}_{4}&\equiv&\kappa_{4}^{\Phi,l}D_{\mu}\Phi D^{\mu}\Phi S^l/M_{Pl}\,.\nn
\eea
In the formulae above\footnote{In the broken phase the impact of these operators may be roughly characterized as: a gauge-kinetic-form altering operator (${\cal O}_{1}$), a Yukawa-altering structure (${\cal O}_{2}$), a gauge-vertex-like correction (${\cal O}_{3}$) and a scalar-kinetic-form altering operator (${\cal O}_{4})$, respectively.} $X_{\mu\nu}$ stands for the gauge field tensor, $f_{i}$ denote matter fermions, $\Phi$ is a generic scalar field, $H^k$ are scalars over which  the SM Higgs doublet is spanned, and $\kappa_{n}$ denote (generally unknown) ${\cal O}(1)$ couplings; for the sake of simplicity the spinorial structure has been suppressed. It is important to notice that not all of these are, however, independent structures from the low-energy effective theory point of view: ${\cal O}_{3}$ and ${\cal O}_{4}$ may be removed from the effective operator basis by use of equations of motion and/or by integrations by parts, see, e.g.,~\cite{Gripaios:2016xuo}. Hence, in what follows, we shall focus entirely on the ${\cal O}_{1}$ and ${\cal O}_{2}$ types of the $d=5$ structures.

%

Despite that, in the broken phase, they usually affect their renormalizable-level couterparts (namely, the gauge-kinetic forms and the Yukawa couplings) at only a relatively small -- order $1\%$ -- level (given by the typical ratio of $\langle S\rangle\sim M_{G}\sim 10^{16}\,$GeV and the Planck scale $M_{Pl}\sim 10^{18}\,$GeV) they can have truly devastating consequences for the robustness of the renormalizable-level results\footnote{Remarkably, both ${\cal O}_{1}$ and ${\cal O}_{2}$ enter the proton lifetime prediction business in more than one way; for instance,  ${\cal O}_{1}$ affects not only the mass of the vector and scalar mediators determined from the gauge unification constraints but also the value of the unified gauge coupling; similarly, ${\cal O}_{2}$ inflicts not only changes in the unitary matrices diagonalizing the masses and, hence, in the BLVN vector currents coupled to the relevant vector mediators, but at the same time, it directly affects also the colour scalar triplet couplings to matter and, hence, the scalar-driven transitions.}, see, e.g.,~\cite{Dixit:1989ff}.
To this end, the most dangerous is ${\cal O}_{1}$  which has been studied thoroughly in many works, see e.g.~\cite{Calmet:2008df,Chakrabortty:2008zk}. Its main effect, i.e., an inhomogeneous shift in the high-scale gauge matching conditions, can inflict a significant shift in the exponent of the functional dependence of the $M_{G}/M_{Z}$ ratio which, even for $1\%$ shifts in the  matching, may change $M_{G}$ by as much as an order of magnitude and, hence, alter $\tau_{p}$ by several orders.    

Concerning ${\cal O}_{2}$, there are many studies in the literature (like, e.g.,~\cite{Ellis:1979fg,Barbieri:1979hc,Dorsner:2005fq}) in which non-renormalizable contributions to the Yukawa couplings  have been added to an originally renormalizable Lagrangian on purpose, usually with the aim to save a renormalizable model suffering from a badly non-realistic Yukawa sector (like in the minimal $SU(5)$ model~\cite{GeorgiGlashow}). Needless to say, this approach is orthogonal to the line of thoughts we want to pursue here, namely, focusing on the potential impact of a-priori unknown Planck-suppressed operators on the existing renormalizable-level predictions, hence, testing their overall robustness. 

In this study, we concentrate on the stability of the tree-level gauge-boson-mediated contributions to the proton decay in the simplest renormalizable unified models based on the $SO(10)$~\cite{FritzschMinkowski}, flipped-$SU(5)$~\cite{GeorgiFlipped, BarrFlipped, DerFlipped} and $SU(5)$~\cite{GeorgiGlashow} gauge groups with respect to several types of uncertainties, either due to the lack of any (or part of the) information about their flavour structure, or due to the presence of only mildly suppressed (up to order $1\%$) Planck-scale induced operators of the  ${\cal O}_{2}$ type above.   

To this end, we first (in Section II) recapitulate the generic analytic observations made by Dorsner and Fileviez-Perez \cite{Dorsner:2004xa,Dorsner:2004jj,FileviezPerez:2004hn} in the realm of the  simplest $SU(5)$, flipped-$SU(5)$ and $SO(10)$ scenarios and complement them with an explicit numerical simulation of the relevant formulae revealing, e.g., an extra room for large cancellation effects in $SO(10)$ GUTs; we will show how these can in some cases boost the uncertainties beyond naive expectation. 
In Section III, we consider a few specific types of renormalizable-level proton lifetime estimates and assess their generic robustness with respect to the effects inflicted by the possible presence of the $M_{Pl}$-induced Yukawa-altering $d=5$ non-renormalizable operators. To this end, we especially focus on a thorough numerical analysis of the stability of the flavour structure of the BLNV currents corresponding to a variety of existing renormalizable-level Yukawa-sector fits~\cite{Joshipura:2011nn,Dueck:2013gca,Altarelli:2013aqa,Meloni:2014rga,Meloni:2016rnt} in the minimal $SO(10)$ and its variants.

\section{Flavour structure of tree-level gauge-mediated amplitudes}\label{secPDecay}
\subsection{Partial decay widths}

 Focusing on tree-level amplitudes mediated by heavy vector bosons arising from terms like \eqref{LagBLNV} in the Lagrangian the relevant partial proton decay widths can be written as\footnote{We assume here that neutrinos are Majorana and some form of a seesaw mechanism is in operation, hence $\nu_R$ is too heavy to be produced in proton decay. 
Should neutrinos be Dirac the sensitivity of the proton widths is less pronounced,  see e.g. \cite{Dorsner:2004xa}.} \cite{proton}
\begin{align}
&\Gamma(p\to \pi^0 e_\beta^+)=C_\pi \,\left\{ \left|c(e_\beta,d^C)\right|^2+\left|c(e_\beta^C,d)\right|^2\right\},\label{gammapi0}\\
&\Gamma(p\to \eta e_\beta^+)=C_\eta\,\left\{ \left|c(e_\beta,d^C)\right|^2+\left|c(e_\beta^C,d)\right|^2\right\},\label{gammaeta}\\
&\Gamma(p\to K^0 e_\beta^+)=C_K \, B_1^2\!\left\{ \left|c(e_\beta,s^C)\right|^2\!+\!\left|c(e_\beta^C,s)\right|^2\right\},\label{gammaK0}
\\
&\Gamma(p\to \pi^+\overline{\nu})=2C_\pi \, \sum_{l=1}^3\left|c(\nu_l,d,d^C)\right|^2\,,\label{gammapi+}\\
&\Gamma(p\to K^+\overline{\nu})= C_K \sum_{l=1}^3\left|B_2\, c(\nu_l,d,s^C)\!+\!B_3\,c(\nu_l,s,d^C)\right|^2\!\!\label{gammaK+}
\end{align}
where incoherent summation over the neutrino flavours is performed, since the neutrinos in the final state are not detected; for similar reasons it is also summed over the chirality of the charged leptons in the final state. The definition of the flavour independent prefactors $C_\pi$, $C_\eta$, $C_K$ and $B_{1,2,3}$ is postponed to Appendix~\ref{appWidths}, let us focus here on the flavour structure of the partial widths determined by the $c$-amplitudes
\begin{align}
&c(e_\alpha,d_\beta^C)=k_1^2(U_C^\dagger U)_{11}(D_C^\dagger E)_{\beta\alpha}+k_2^2(D_C^\dagger U)_{\beta 1}(U_C^\dagger E)_{1\alpha},\label{cedC}\\
&c(e_\alpha^C,d_\beta)=k_1^2[(U_C^\dagger U)_{11}(E_C^\dagger D)_{\alpha\beta} +(U_C^\dagger D)_{1\beta}(E_C^\dagger U)_{\alpha 1}],\label{ceCd}\\
&c(\nu_l,d_\alpha,d_\beta^C)\!=\!k_1^2(U_C^\dagger D)_{1\alpha}(D_C^\dagger N)_{\beta l}\!+\!k_2^2(D_C^\dagger D)_{\beta\alpha}(U_C^\dagger N)_{1 l},\label{cnudd}
\end{align}
where $k_i=g_{G}/(\sqrt{2}M_{i})$ with $g_{G}$ denoting the universal gauge coupling at the GUT scale and $M_{1,2}$ encoding the masses of the heavy vectors with the  $SU(3)_c\times SU(2)_L\times U(1)_Y$ quantum numbers $(\mathbf{3},\mathbf{2},5/6)$ and $(\mathbf{3},\mathbf{2},-1/6)$, respectively. In the flipped-$SU(5)$ scenario only the latter is present and, hence, $k_1=0$; similarly, $k_2=0$ in case of the ordinary $SU(5)$. In the $SO(10)$ GUTs both $k_{1,2}$ are non-zero. The unitary rotations entering the coefficients \eqref{cedC}-\eqref{cnudd} are defined as
\begin{align}
Y_d^{\mathrm{diag}}&=D_C^T Y_{d} D\,, \quad Y_u^{\mathrm{diag}}=U_C^T Y_{u} U\,,
\label{rotace}  \\
\nonumber Y_e^{\mathrm{diag}}&=E_C^T Y_{e} E\,,\quad\,
Y_{\nu}^{\mathrm{diag}}=N_C^T Y_{\nu} N\nn,
\end{align}
or, generically, $Y_f^{\mathrm{diag}}=F_C^T Y_{f} F$, 
where $Y_f$ are the relevant effective SM Yukawa matrices (in the RL basis) which, in their diagonal form $Y_f^{\mathrm{diag}}$ (and after multiplication by the electroweak VEV),  yield the physical masses of the fermions of type $f$ at 
 $M_{G}$. Let us note that if some of the mass matrices happen to be symmetric (as, e.g., in case of Majorana neutrinos), then the LH and RH rotations are identical.

Up to a possible multiplication by phase factors the LH rotations in \eqref{cedC}-\eqref{cnudd} are correlated to the physical CKM and PMNS matrices via
\bea\label{CKM}
 V_{CKM}&=&K_1 U^\dagger D K_2\equiv K_1  \tilde{V}_{CKM}K_2  \,,\\ \label{PMNS}
 V_{PMNS}&=& K_3 E^\dagger N K_4\equiv K_3  \tilde V_{PMNS} K_4\,.
\eea
where the tilded quantities correspond to their ``raw'' form, i.e., the form before the freedom in the phase redefinition of the fermionic fields\footnote{Here $K_1$ and $K_3$ contain 3 phases, $K_2$ contains 2 phases and $K_4$ is a unit matrix in case of Majorana neutrinos or contains up to 2 phases if they are Dirac.} has been exploited.
Note that, in general, these are also the \emph{only} experimental constraints on the rotations in (\ref{rotace}) one has; without extra information about the flavour structure of a particular model $U_C$, $D_C$ and $E_C$ (assuming Majorana $\nu$'s) are completely free. 

\subsection{Model independent constraints}\label{secYukGen}
With this information at hand, number of semi-analytic observations about the relative sizes of the uncertainties plaguing the partial widths (\ref{gammapi0})-(\ref{gammaK+}) due to the lack of grip on most of the flavour structures therein has been made~\cite{Dorsner:2004xa, Dorsner:2004jj} even without any extra model-dependent assumptions on the shape of the mixing matrices~in (\ref{rotace}). In particular, the question whether the total proton decay width\footnote{Here it is implicitly assumed that the rates into final states with higher spin mesons will be suppressed; moreover, their flavour structure is essentially the same as for the spin zero modes (\ref{gammapi0})-(\ref{gammaK+}) and, hence, qualitative changes of the results are not expected.} can be zeroed out, i.e., whether the proton decay can be ``rotated away'' was addressed. 

Remarkably enough, in case of the flipped $SU(5)$ unifications where $k_1=0$, $k_2\neq 0$ all the amplitudes \eqref{cedC}-\eqref{cnudd} can be indeed pushed to zero \cite{Dorsner:2004jj} if one arranges for $(U_C^\dagger E)_{1\alpha} = 0$ for $\alpha=1,2$ and $(D_C^\dagger D)_{\beta\alpha}=0$ for the combinations $\{\beta,\alpha\}=\{1,1\},\{1,2\},\{2,1\}$. This can be done easily if there is no correlation among the LH and RH rotations which, however, may not be the case in the most minimal models, see below.

\begin{figure}
\begin{center}
\includegraphics[width=\columnwidth]{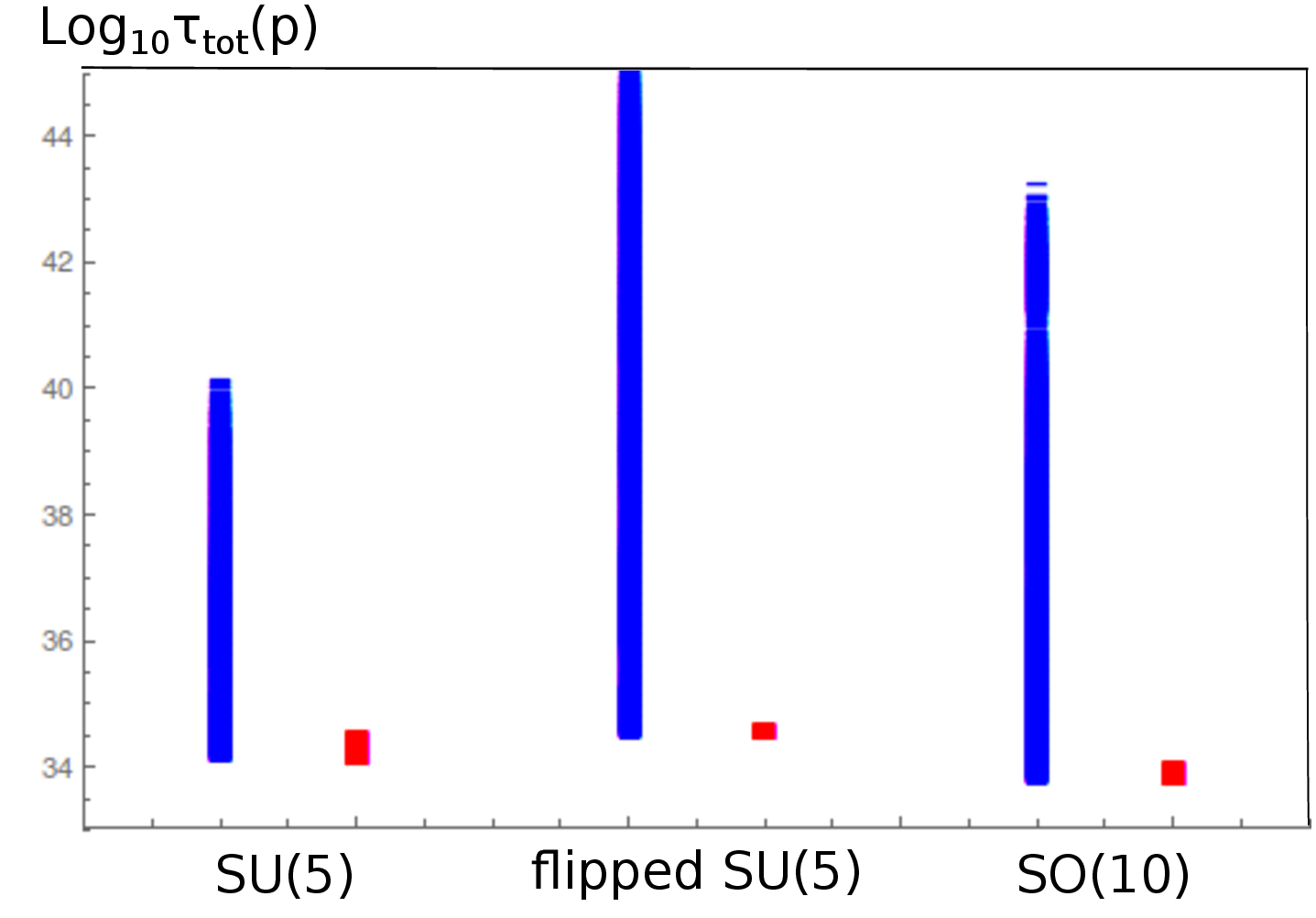}
\caption{Total proton lifetime for different choices of the flavour-dependent coefficients \eqref{cedC}-\eqref{cnudd} is displayed for different gauge groups (with $M_{1,2}\sim 0.3\times 10^{16}$ GeV generally assumed in order for the current SK bound on the partial width in the $p\to\pi^0 e^+$ channel to be saturated in the $SO(10)$ case, see also FIG.~\ref{figGold}). The red bars correspond to ``minimal'' renormalizable scenarios which, in all cases of our interest, feature extra correlations among the relevant mixings due to, e.g., symmetry of the underlying Yukawa matrices in certain sectors, see Section \ref{secYukRen}. In case of the blue bars, no such correlations are assumed.}
\label{figRot}
\end{center}
\end{figure}

On the contrary, in the standard $SU(5)$ GUTs where $k_1 \neq 0$, $k_2 = 0$ the non-zero value of $|(V_{CKM})_{13}|$ element forbids to rotate the proton decay away \cite{Dorsner:2004jj}; however, the small size of this parameter admits up to about $\O(10^{-3})$ suppression of the amplitudes~\eqref{cedC}-\eqref{cnudd} and, hence, up to some $\O(10^{-6})$ suppression of the total decay width (see FIG.~\ref{figRot}). Consequently, the  proton decay can be ``hidden'' from the current experiments even if the unification scale would be as low as $10^{14}\,\GeV$ \cite{Dorsner:2004xa}.

In case of $SO(10)$ unifications with $k_1\neq 0$, $k_2\neq 0$ one would naively expect similar lower bounds on the amplitudes~\eqref{cedC}-\eqref{cnudd} which, in turn, might suggest the same $\O(10^{-6})$ maximum suppression of the total proton decay width. However, with both $k_1$ and $k_2$ at play, destructive interference effects may sometimes occur in all the coefficients \eqref{cedC}-\eqref{cnudd} which would make the total proton decay width even smaller than that, see FIG.~\ref{figRot} for a numerical simulation (with $M_1=M_2$ assumed for simplicity\footnote{This is justified by the fact that for $M_{1} \ll M_{2}$ or $M_{1} \gg M_{2}$ one of the two previously discussed cases is effectively recovered.}). 

\subsection{Minimal renormalizable settings}\label{secYukRen}
In scenarios in which the scalar sector is specified, 
extra correlations among the flavour rotations (\ref{rotace}) are often in operation. If, for instance, some of the Yukawa matrices happen to be symmetric the RH and LH rotations are strongly correlated and significant simplifications may occur in formulae \eqref{cedC}-\eqref{cnudd}.  

Namely, in the original Georgi-Glashow $SU(5)$ model \cite{GeorgiGlashow}, the symmetry of the fermionic $10\otimes 10$ bilinear in the flavour space implies that the up-quark Yukawa matrix is symmetric and, hence, $U=U_C$ . In this case the partial widths with (unidentified) neutrinos in the final state become entirely driven by the CKM matrix elements \cite{Dorsner:2004xx} and the uncertainty in the total proton lifetime shrinks considerably as depicted in FIG.~\ref{figRot}.
On the other hand, the basic $SU(5)$ flavour structure considered above should be extended in order to deal with the down-type-quark--charged-lepton degeneracy issues (and other notorious problems of the simplest $SU(5)$ unifications concerning, e.g., the unification of gauge couplings or non-zero neutrino masses). 
New fields (such as 45-dimensional scalar representation \cite{Georgi:1979df,Dorsner:2006dj}) and/or higher-dimensional operators \cite{Ellis:1979fg,Dorsner:2005fq, Dorsner:2006hw} are usually employed for that sake. In either case the exact symmetry of $Y_u$ is lifted. For this reason, the red bar for $SU(5)$ unifications in FIG.~\ref{figRot} is to be taken as purely illustrative.

In the flipped $SU(5)$ scenario, the RH quark field $d^{C}$ is swapped with $u^{C}$  and, hence, it is the down-type Yukawa that gets symmetric in the minimal settings which implies $D=D_C$. This leads~\cite{Dorsner:2004xx} to very simple relations
\begin{align}
\label{symFlipped}&\Gamma(p\to \pi^+\overline{\nu})=2C_\pi\, k_2^4\,,\\
\nonumber &\Gamma(p\to K^+\overline{\nu})= 0\,,
\end{align}
and, consequently, to a significant reduction of the uncertainty in the proton lifetime estimates
, see FIG.~\ref{figRot}. Let us emphasize that in case of flipped $SU(5)$ the constraint $D=D_C$ is satisfied also by fully realistic models including all necessary ingredients like, e.g., non-zero neutrino masses \cite{Das:2005eb,Rodriguez:2013rma}.

In case of the renormalizable $SO(10)$ unifications, the minimal potentially realistic choice of the scalar fields shaping the Yukawa sector corresponds to a 10-dimensional vector and a 126-dimensional 5-index antisymmetric self-dual tensor. Both these yield symmetric Yukawa couplings and, thus, \emph{all} RH rotations in (\ref{rotace}) are strongly correlated with the LH ones. This leads to \cite{FileviezPerez:2004hn} 
\begin{align*}
&\Gamma(p\to \pi^+\overline{\nu})=2 C_\pi\times \\
&\times \left\{k_1^4 \left|(V_{CKM})_{11}\right|^2 + k_2^4 + 2 k_1^2 k_2^2 \left|(V_{CKM})_{11}\right|^2\right\}\, ,\\
&\Gamma(p\to K^+\overline{\nu})= C_K\times\\
&\times k_1^4\left(B_2^2 \left|(V_{CKM})_{11}\right|^2 + B_3^2 \left|(V_{CKM})_{12}\right|^2\right)
\end{align*}
Again, the uncertainty in the total proton lifetime shrinks enormously, see FIG.~\ref{figRot}. 

\subsection{Two-body p-decay amplitudes with a charged lepton in the final state}
Unlike for the (anti)neutrino channels above the amplitudes of the two-body partial proton decay widths with a charged lepton in the final state are generally driven by non-trivial combinations of the mass-diagonalization matrices~\eqref{rotace} with only an indirect\footnote{Sometimes such a connection may not even be made at all - a classical example would be the lack of constraints for the flavour structure of the RH  leptoquark currents in models with Yukawa couplings featuring no extra symmetries.} connection to the low-energy flavour observables.
   
Hence, in order to get any theoretical grip on these channels, one must resort to a specific model and construct a detailed map of all possible phenomenology-compatible flavour patterns, i.e., a complete set $\{F,F_C\}$ with $F=U,D,E,N$ defined in~\eqref{rotace}. 
Technically, this information can be obtained from a thorough analysis of the fits of the Yukawa structure underlying the quark and lepton mass matrices $M_{f}$. 
For each such setting the relevant amplitudes can then be fully reconstructed and, if desired, extremized over the entire set of such configurations. 

This, however, is a highly non-linear game and, thus, in most cases~\cite{Joshipura:2011nn,Dueck:2013gca,Altarelli:2013aqa,Meloni:2014rga,Meloni:2016rnt}, the authors resort to the renormalizable-level approximation in which one typically deals with a limited number $n$ of independent Yukawa matrices $Y^{k}$ of  the $d=4$ Yukawa lagrangian entering the renormalizable-level matching conditions for the effective low-energy couplings $\tilde{Y}_{f}$ in the form (no summation over $f=u,d,\ell,\nu$, generation indices suppressed):
\be\label{renfits}
\tilde{Y}_{f} = \frac{1}{v}\sum_{k\in D_{4}}c_{f}^{k}v^{k}_{f}Y^{k}\equiv Y_f\,.
\ee
Here $v$ denotes the electroweak VEV serving merely as a normalization factor, $v_{f}^{k}$ stand for the projections of the SM Higgs VEV onto the underlying-theory doublets $H^k$ that couple to the fermionic bilinears at the $d=4$ level and $c^{k}_f$ cover all remaining constant ${\cal O}(1)$ numerical factors (Clebsches, symmetry coefficients and so on). It is clear that if $D_{4}$, the set of indices corresponding to the doublets relevant at the $d=4$ level, is small (i.e., if the number of such doublets is less than 4) the fits of $Y_{f}$'s in terms\footnote{The fundamental doublet projections $v_{f}^{i}$ are, in principle, calculable functions of the scalar potential parameters and, usually, turn out to be correlated among themselves. An extreme example of this is the situation in the minimal SUSY $SO(10)$ model~\cite{Aulakh:2003kg} which was eventually discarded~\cite{Bertolini:2006pe,Aulakh:2005mw} just due to such correlations.} of $Y^{i}$'s may be quite non-trivial, $\{F,F_C\}$ strongly constrained and, thus, the theory predictive.  
\section{d=5 Planck-scale flavour effects}
\label{secPlanck}
In reality, however, the renormalizable-level fits may be incomplete because the effective Yukawa matrices may be affected by physics at the Planck-scale $M_{Pl}$ which, in the effective theory picture, may enter the game by means of non-renormalizable $d>4$  operators. For instance, the presence of the $d=5$ operators of the Yukawa type [class ${\cal O}_{2}$ in the list (\ref{OpsList})] inflicts additional shifts in the relevant matching conditions between the effective (running) low-energy Yukawa couplings $Y_f$ and the underlying GUT-theory couplings 
in the form 
\be\label{sumrule}
\tilde{Y}_f=Y_f\;\; (\text{at }d=4)\quad \to \quad \tilde{Y}_f=Y_f+\Delta Y_f\;\; (\text{at }d>4),
\ee
where $\Delta Y_f$ may be schematically written as (again, no summation over $f$ and no generation indices)
\be\label{nonrenY}
 \Delta Y_f v \equiv \sum_{k\in D_{5}}\sum_{l\in S_{5}}\kappa^{kl}c_{f}^{kl}v^{k}_{f}\frac{\vev{S^{l}}}{M_{Pl}}\,.
\ee
Here the meaning of $v_{f}^{k}$ is the same like above, $D_5$ is the set of relevant doublet indices that is summed over (note, however, that $D_{5}$ is not\footnote{At $d=5$ there may be contributions in (\ref{nonrenY}) from scalar multiplets containing SM doublets that would not be present in (\ref{renfits}) and vice versa. As an example consider the $126$ and $210$ scalars in $SO(10)$ - the latter irrep can, indeed, couple to the fermionic bilinears only at the $d=5$ level.} necessarily the same as $D_{4}$ in (\ref{renfits})), $S_5$ indexes the SM singlets with GUT symmetry breaking VEVs, $\kappa^{kl}$ are the coefficients of the relevant $d=5$ operators as in (\ref{OpsList}) and, as before, $c^{kl}_f$ cover all the remaining numerical factors.

Comparing (\ref{renfits}) with \eqref{nonrenY} and assuming that the coefficients $c_{f}^{kl}$ and $\kappa^{kl}$ are at most $\mathcal{O}(1)$ one may expect that the Planck-induced contributions $\Delta Y_f$ to the full effective Yukawa coupling $\tilde{Y}_f$ in (\ref{sumrule}) should typically come with an additional 
$
M_G/M_{Pl} \sim \mathcal{O}(10^{-2})
$
factor.

Although such corrections may naively appear to be small, they may make the matrices diagonalising the complete $\tilde{Y}_f$'s (which we shall from now on denote by $\{\tilde{F},\tilde{F}_C\}$) significantly different from those obtained at the $d=4$ level (denoted by $\{{F},{F}_C\}$). The point is that the entries of the $Y_f$ matrices (and, most importantly, their eigenvalues) are often smaller than $\mathcal{O}(M_G/M_{Pl})\sim \mathcal{O}(10^{-2})$; hence, even an $\mathcal{O}(10^{-2})$ correction can be enough to change the shape of the ``small'' part of the Yukawa matrix completely (we further elaborate on the formal aspects of this in Appendix \ref{appMath}). 

On the other hand, even though the two sets of rotation matrices $\{{F},{F}_C\}$ and $\{\tilde{F},\tilde{F}_C\}$ may look dramatically different, the resulting partial proton decay widths may still be rather similar since the relevant formulae \eqref{gammapi0}-\eqref{gammaK+} depend only on their {\em specific products}  and, as we shall see, in some cases there may be reasons to expect significant cancellations of such $d>4$ effects.

\subsection{Robustness of the renormalizable-level $p$-decay  estimates with respect to $d>4$ effects}

At first glance, it may seem rather hopeless to attempt to say anything general enough to be interesting about the possible differences of the two sets of matrices $\{{F},{F}_C\}$ and $\{\tilde{F},\tilde{F}_C\}$ -- i) either one fits the effective Yukawa matrices in a renormalizable model and then has little or no grip onto a typically yet larger set of the higher-order operators, or ii) one includes non-renormalizable operators into the game right away (because it may be necessary to do so otherwise no consistent parameter-space points may be found at all, see e.g.~\cite{Bajc:2002pg,Dorsner:2006hw,Azatov:2008vu,Ajaib:2013zha}
) and then, naturally, never asks about the renormalizable case because it makes little sense. 

The main scope of this work is to argue that the situation corresponding to the case\footnote{For obvious reasons we do not intend to elaborate on case ii) here.} i)  above may be slightly more subtle and, in fact, under some circumstances, one may say something sensible about the robustness of the p-decay estimates based on the renormalizable-level Yukawa sector fits even without a detailed knowledge of the structure of the higher-dimensional contributions therein.
\subsubsection{First look: The problem in full generality\label{fullexercise}}
Let us start with assuming for the moment an ideal world in which we have enough computing power to generate all (with perhaps some given granularity in practice) fits of the effective Yukawa matrices $\tilde{Y}_f$ in terms of the underlying renormalizable-level Yukawa couplings subject to sum-rules dictated by the unified model under scrutiny and perhaps even more power to repeat the same exercise for the more complicated $d>4$ case. 

The former, in other words, amounts to getting first all possible structures of $Y^k$'s in (\ref{renfits}) associated to all attainable configurations of $v^l$'s which, in the $d=4$ case, yield the $\tilde{Y}_f$'s that (after the necessary renormalization group evolution to our energies) encode the desired spectra of the SM fermions together with their mixing in the charged-current interactions (aka the CKM and PMNS matrices). With these at hand one would then easily derive the complete set of the possible $\{F,F_C\}$ matrices which shall be eventually used to estimate the proton lifetime {\em and, in particular, the associated theoretical uncertainty corresponding to the fact that the low energy data can not pinpoint the ``true" solution among all these possibilities}. In the second step one repeats the same exercise with just a little bit more of freedom due to the presence of the extra couplings corresponding to the $d>4$ operators and derives all possible $\{\tilde{F},\tilde{F}_C\}$ associated to these ``extended" fits together with the relevant proton lifetime estimates.

Given this it is immediately clear that:
\begin{itemize}
\item
The set $\cal F$ of all thus obtained ``realistic" $\{F,F_C\}$ is a subset of the set  $\tilde{\cal F}$ of all ``realistic" $\{\tilde{F},\tilde{F}_C\}$;
\item
Without any specific constraint on the size of the higher-order operators the set of the ``realistic'' $\{\tilde{F},\tilde{F}_C\}$ may be so large that one effectively looses any grip on the vector leptoquark interactions and, subsequently, the theoretical uncertainties of the proton lifetime estimates within such scenarios rocket (and, hence, exhibit the behaviour depicted by the blue bars in FIG.~\ref{figRot}).     
\end{itemize}
The point we will try to make is that, in some cases, even a simple extra assumption such as an additional ${\cal O}(10^{-2})$ suppression associated to each subsequent step on the effective operator ladder, in conjunction with specific features of the renormalizable-level fits such as their symmetry in the generation space,  may be enough to correlate $\tilde{\cal F}$ to ${\cal F}$ to such a degree that the proton lifetime estimates obtained within the humble $d=4$ approach may actually represent a very good approximation to the ``true'' (i.e., full theory) predictions.  

\subsubsection{The trick: Small perturbations and continuity\label{sect:continuity}}
Needless to say, the programme sketched in the previous part (i.e., obtaining the sets ${\cal F}$ and $\tilde{\cal F}$ -- {\em both complete} -- and comparing the spans of the associated proton lifetime estimates in order to asses the robustness of those based only on ${\cal F}$) is intractable\footnote{Besides intractability it does not even make sense to do that because with the complete  $\tilde{\cal F}$ at hand nobody would care about ${\cal F}$ anymore.}.  However, for small $|\Delta Y_f|\lesssim 10^{-2}$ 
(for all $f$'s) the task to learn something about $\tilde{\cal F}$ can be accomplished even without  embarking on its full determination by  {\em assuming continuity} in the change of the fitted renormalizable-level Yukawa couplings as functions of the size of the non-renormalizable contributions. 

Technically, what we have in mind is that the full $\tilde{\cal F}$ set obtained upon fitting the complete non-renormalizable structure $Y_f+\Delta Y_f$ with no assumptions made on $Y_f$ and $\Delta Y_f$ (besides the smallness of the latter, see~(\ref{sumrule})), will be essentially\footnote{One may argue that in this way we are  mapping the $\tilde{\cal F}$ set corresponding to $2\varepsilon$ rather than the original $\varepsilon$ size of the $d>4$ perturbations. This is true but, at the same time, we do not really care because in the semi-qualitative discussion to follow this makes no significant difference.} the same as the set $\tilde{\cal F}'$ obtained by fitting a slight variation of the original formula, namely $Y_f+\Delta Y_f \to Y'_f+\Delta Y_f'\equiv \tilde{Y}_f'$ where $Y_f'$ {\em is assumed to run over the set of all good fits of the renormalizable-level case only}.

The point is that due to continuity assumption the choice of specific $Y_f'$'s instead of a fully general $Y_f$ inflicts only a small change in the structure from which $\tilde{\cal F}$ would have been derived. Indeed,  this change can be modelled by a mere reshuffling of the set of the Planck-scale-induced corrections that would have to be summed over anyway to get the complete $\tilde{\cal F}$ (indeed, $\Delta Y_f'=\Delta Y_f+Y_f-Y_f'$); in this respect, the $Y_f\to Y_f'$ and $\Delta Y_f\to \Delta Y_f'$ replacements qualitatively correspond to a mere choice of a specific reference element in the set of all possible $d>4$ contributions\footnote{One may wonder if the set ${\cal \tilde{F}}'$ is not subject to an extra restriction compared to ${\cal \tilde{F}}$ due to the fact that $Y_f'$ and $\tilde{Y}_f' = Y_f'+\Delta Y_f'$ must have the same generalized eigenvalues in order to yield correct fermion masses. However, the Lemma~\ref{pokus} in Appendix~\ref{appMath} suggests that arranging for the correct generalized eigenvalues (up to the order of ${\cal O} (\eps^2)$) does not restrict the set $\{\tilde{F}',\tilde{F}'_C\}$ at all.}!
Needless to say, this leads to an enormous simplification 
of the general problem described in Sect.~\ref{fullexercise} and, as such, it represents {\em the central point of this study} (and, in fact, the very key to its practical feasibility).
Hence, we may simplify our life in mapping the $\tilde{\cal F}$ set by reformulating the general exercise described in Sect.~\ref{fullexercise} into a much more tractable one of employing only the {\em specific} shapes of the renormalizable-level Yukawa couplings corresponding to the renormalizable-level fits whilst keeping {\em fully general} (i.e., unspecified but small) only the $d>4$  contributions.  Thus obtained $\tilde{\cal F}'$ should be essentially the same as $\tilde{\cal F}$.    

\vskip 3mm
In conclusion, what one should do in practice is to take all possible renormalizable-level fits $Y_f'$ of the effective Yukawa structure of the given model; then, for each such $Y_f'$, construct the sums 
\be\label{central}
\tilde{Y}_f'= Y'_f+\Delta Y_f'
\ee 
with all $\Delta Y_f'$'s obeying  $|\Delta Y_f'|\lesssim \mathcal{O}(M_G/M_{Pl}) \sim \mathcal{O}(10^{-2})$, take all cases in which $\tilde{Y}_f'$'s happen to give the right SM fermion masses and mixings (as $Y'_f$ do by construction) and, eventually, derive and save the corresponding $\{\tilde{F}',\tilde{F}'_C\}$'s. These will be, subsequently, used as inputs of a refined $p$-decay analysis in the models of interest.  
\subsubsection{Further comments\label{secComentsMethod}}
There are perhaps few more comments worth making here: 
First, it may well be the case that the complete set of {\em all possible} renormalizable-level fits that the trick above relies on may not be fully available as its determination represents a formidable task on its own. To this end, in what follows we shall do what we can, i.e., we shall take a look onto just a specific (and small) set of popular and simple enough scenarios and, within these, confine ourselves to {\em all available} sets of $Y_f'$'s that may be found in the corresponding literature. In this sense, the results presented in the next section may not be completely general; nevertheless, they are not useless as they admit to estimate the robustness of at least the existing proton lifetime calculations based on the renormalizable-level flavour fits.       

Second, the entries of the individual $\Delta Y_f$'s of eq.~(\ref{nonrenY}) may be in general further restricted due to the extra symmetries of the underlying effective operators in the generation space. 
 Out of the restrictions of this kind, the case with both $Y_f$ and $\Delta Y_f$ symmetric for certain $f$'s is of most interest since, in such a case, also the effective Yukawa couplings of Eq.~\eqref{sumrule} inherit this symmetry. Consequently,  the set $\{\tilde{F}',\tilde{F}'_C\}$ has the simplified structure described in Section~\ref{secYukRen} and the total proton lifetime uncertainties are contained within the red bars in FIG.~\ref{figRot}. Remarkably, this is the case for the $SO(10)$ GUT featuring the 10- and 126-dimensional Yukawa active scalars with the first-stage gauge symmetry breaking driven by the 54-dimensional scalar\footnote{Indeed, in $SO(10)$ one has $10\otimes 54 = 10\oplus 210 \oplus 320$ and $126\otimes 54 = \overline{126} \oplus 1728 \oplus 4950$ and the only representations which can be contracted to form a singlet at $d=5$ with two 16-dimensional fermion representations in these sums are $10$ and $\overline{126}$; this then yields symmetric Yukawa couplings. The Planck-suppressed operators of the type $\mathcal{O}_2$ in \eqref{OpsList} with $H=10/126$ and $S=54$ would then imply that all $\Delta Y_f$'s are also symmetric.} (see, e.g., \cite{Babu:2015bna} for a recent study including the Yukawa sector fits). Such situation is, however, rather exceptional. For instance, if the $SO(10)$ gauge group was broken by 45 instead, $\Delta Y_f$'s would contain also an antisymmetric part due to the presence of the 120-dimensional representation in the product $45\otimes 10$. As already mentioned in Section~\ref{secYukRen}, the situation is similar in the case of the simplest $SU(5)$ unifications, where the $d=5$ non-renormalizable operators destroy the symmetry of the up-type quark Yukawa matrix (see, e.g., \cite{Dorsner:2006hw}). 
Thus, in what follows, we decide not to impose any extra generation-space symmetries onto the $\Delta Y_f$'s of eq.~(\ref{nonrenY}).

Third, the $\Delta Y_f$'s of eq.~(\ref{nonrenY}) may be, in principle, further correlated across different flavours (i.e., $f$'s) due to their common origin from a potentially limited set of available $d>4$ effective operators. Such correlations are, however, strongly model-dependent; therefore, we choose to ignore such nuances in the current analysis. This means that our results may be viewed as corresponding to the most pessimistic situation and, in reality, the uncertainties of the proton decay estimates within specific scenarios may be smaller. 
If, on the other hand, the partial proton decay widths turn out to exhibit a certain degree of robustness with respect to the uncorrelated $\O(M_G/M_{Pl})$ perturbations, the same behaviour should be reflected also in the real, i.e., more constrained case. 

In this respect, the approach of imposing no extra constraints onto the shapes of $\Delta Y_f'$'s (besides their smallness) is perhaps the only strategy which can, on one hand,  reflect the specifics of the underlying renormalizable-level fits and, at the same time, save thus obtained results from any further model-dependent assumptions.

\subsubsection{The numerical approach}
Let us now describe the technical aspects of the numerical analysis of formulae~(\ref{central}).

Since both $Y_f'$ and $\tilde{Y}'_f$ are assumed to yield the same physical masses (see Section~\ref{sect:continuity}), one finds
$$Y^{\mathrm{diag}}_f=F_C^{\prime T} Y_f' F' = \tilde{F}_C^{\prime T} \tilde{Y}_f' \tilde{F}'.$$
The perturbation $\Delta Y_f'$ can be, hence, expressed in terms of the varied rotation matrices $\tilde{F}'$ and $\tilde{F}_C'$. Note that it is more convenient to search through the space of the unitary matrices $\tilde{F}'$ and $\tilde{F}_C'$ instead of the  perturbations $\Delta Y_f'$ since then the constraints regarding the CKM and PMNS matrices \eqref{CKM}-\eqref{PMNS} can be easily implemented. Consequently, our strategy is to exploit the set of all possible shapes of $\tilde{F}'$ and $\tilde{F}_C'$ satisfying \eqref{CKM}-\eqref{PMNS} and to check only subsequently whether the resulting perturbation of the Yukawa matrix is as small as required, i.e., whether
\begin{equation}\label{deltaYf}
|\Delta Y_f'|=|\tilde{F}^{\prime *}_{C}{Y}^{\mathrm{diag}}_f\tilde{F}^{\prime\dagger}-Y_f'|\lesssim 10^{-2}
\end{equation}
is satisfied for $f=u,d,e$.\footnote{We constrain the matrix $N$ in \eqref{rotace} only by the PMNS matrix relation \eqref{PMNS} since $Y_\nu$ is usually computed from different Yukawa couplings according to some type of seesaw mechanism, and the corresponding constraints on $\Delta Y_\nu$ would be more complicated and model dependent. This follows our strategy to consider the ``worst case'' scenario; in reality, the true uncertainties in the corresponding proton lifetime may be more constrained. Moreover, since the channels with neutrinos in the final state are always incoherently summed over in \eqref{gammapi0}-\eqref{gammaK+}, we do not expect that further constraining $N$ would have any significant effect.}

Finally, let us comment on the choice of the renormalizable fits that serve as $Y_f'$'s in \eqref{central}. In case of the $SO(10)$ unifications, exact shapes of the fitted Yukawa matrices are available in the literature \cite{Joshipura:2011nn,Dueck:2013gca,Altarelli:2013aqa,Meloni:2014rga,Meloni:2016rnt}; these served as inputs for $Y_f'$'s in our numerical analysis. 

On the other hand, to our best knowledge, neither for the $SU(5)$ nor the flipped $SU(5)$ models any reasonably exhaustive classification of working fits of Yukawa sectors of the minimal potentially realistic and renormalizable scenarios is available in the literature. 
Nevertheless, it may still be interesting to check how much the set $\{\tilde{F}',\tilde{F}'_C\}$ varies from $\{F',F_C'\}$ corresponding to a set of essentially random choices of $Y_f'$'s which, however, are still assumed to respects at least the basic symmetry properties inherent to the minimal models\footnote{In practice, the relations \eqref{rotace} were used, i.e., the diagonal part $Y_f^{\mathrm{diag}}$ was inferred from the Yukawa coupling running in the given model and random $\tilde{F}'$ and $\tilde{F}_C'$ satisfying the constraints on CKM and PMNS matrices \eqref{CKM}-\eqref{PMNS} were chosen.}, see Section~\ref{secYukRen}.  
In case of the flipped $SU(5)$, symmetric $Y_d'$ was always assumed for the starting point while in case of the ordinary $SU(5)$ both symmmetric and non-symmetric versions of $Y_u'$ were checked since, in realistic models, the latter options is usually realized as explained in Section~\ref{secYukRen}.

\subsection{Results\label{sec:results}}
\subsubsection{Simplest SO(10) GUTs.}

\begin{figure}
\begin{center}
\includegraphics[width=\columnwidth]{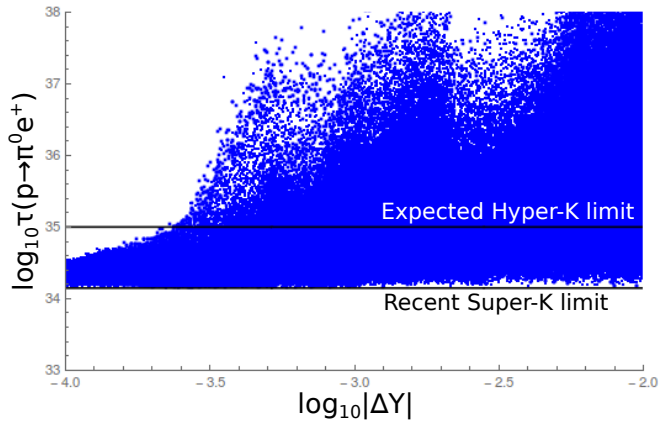}
\caption{The $p\to \pi^0 e^+$ partial proton lifetime in case of the $SO(10)$ unifications as a function of the size of the Planck-induced perturbations $|\Delta Y|$ \eqref{DeltaY}. The renormalizable-level setting of the Yukawa matrices $Y_f'$ corresponds to the fit explicitly given in \cite{Dueck:2013gca} where the $10_H\oplus 126_H$ Higgs content and the normal neutrino mass hierarchy are assumed.
Similar behaviour with large spread of the partial proton lifetime values was obtained also for other individual decay channels and also when other fits in \cite{Dueck:2013gca} served as the initial point.}
\label{figGold}
\end{center}
\end{figure}

\begin{figure}
\begin{center}
\includegraphics[width=\columnwidth]{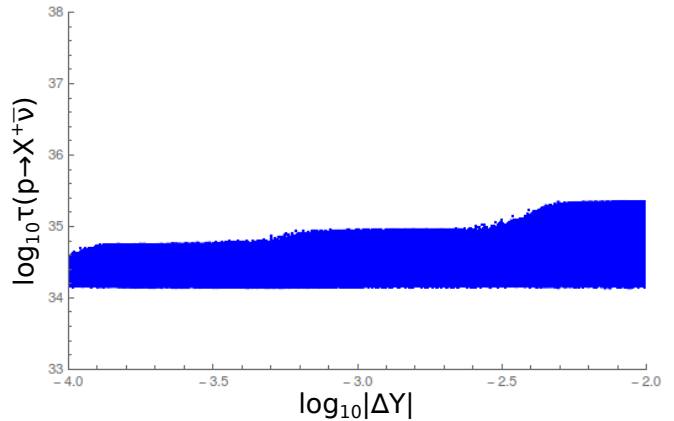}
\caption{The inverse of the sum of the partial $p\to X^+\overline{\nu}$ decay widths for $X=\pi$, $K$ in case of the SO(10) unifications (the same renormalizable-level point as in FIG.~\ref{figGold} was used, and again similar behaviour was observed also for other initial settings based on fits from \cite{Dueck:2013gca}). Note, however, that the behaviour of the \emph{individual} neutrino-final-state channels is similar to the situation for the golden channel (see FIG.~\ref{figGold}).}
\label{figSO10-nus}
\end{center}
\end{figure}

\begin{figure}
\begin{center}
\includegraphics[width=\columnwidth]{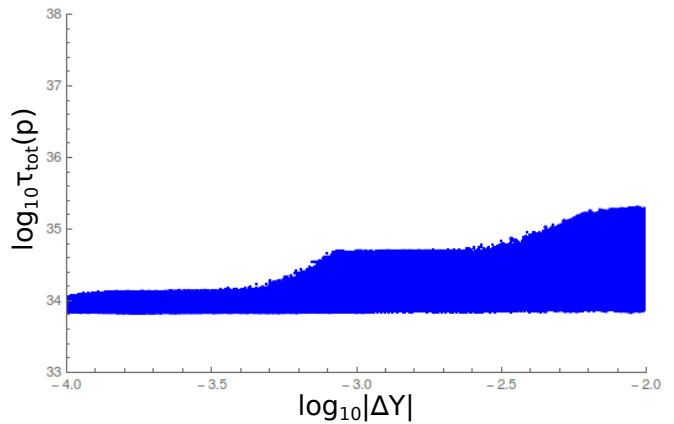}
\caption{The total proton lifetime in case of the minimal SO(10) unification. 
}
\label{figSO10-total}
\end{center}
\end{figure}

Given their relatively rigid Yukawa structure, the $SO(10)$ GUTs provide an ideal setting for us here since a decent number of renormalizable-level Yukawa fits available in the literature 
can serve as a starting point for our numerical analysis. In total, 8 different Yukawa sector fits for non-supersymmetric $SO(10)$ models available in \cite{Joshipura:2011nn,Dueck:2013gca,Altarelli:2013aqa,Meloni:2014rga,Meloni:2016rnt} was studied, including both the cases when $10_H\oplus 126_H$ `Yukawa-active' Higgs fields have been taken into account, hence, the renormalizable-level mass matrices were symmetric, and when also the antisymmetric contribution due to the presence of $120_H$ was added; no significant qualitative differences were observed. As an example, the scan over the space of possible Yukawa matrix perturbations based on the fit obtained in \cite{Dueck:2013gca} considering $10_H\oplus 126_H$ Higgs sector and the normal neutrino mass hierarchy is presented in plots in FIGs.~\ref{figGold} - \ref{figSO10-total}.

When computing the partial proton decay widths, $M_{1}=M_{2}\sim 0.3\times 10^{16}$ GeV was fixed for which the current SK bound on the proton lifetime in the ``golden'' channel $p\to\pi^0 e^+$ is just saturated. The uncertainty in the (inverse) partial proton decay widths or their sums was then plotted against the maximum size of the perturbations \eqref{deltaYf}
\begin{equation}\label{DeltaY}
|\Delta Y| \equiv \max_{f=u,d,e} |\Delta Y_f'|.
\end{equation}

The spread in the individual (inverse) partial proton decay widths was observed to be large for a wide range of $|\Delta Y| \lesssim 10^{-2}$, see FIG.~\ref{figGold} for the example of the ``golden'' channel $p\to \pi^0 e^+$. This, unfortunately, means that even with a fit to Yukawa sector at hand, robust predictions for the individual decay channels are in general impossible. The same behavior with large uncertainties was observed also if it was summed over all the partial widths with the charged leptons in the final state. On the other hand, the situation became much more favourable when the neutrino channels were summed over as shown in FIG.~\ref{figSO10-nus}. Let us note that this behaviour can be understood recalling that it is summed over the neutrino species in the final state for the neutrino channels, whereas the production of the $\tau$ lepton is kinematically forbidden, hence, there is more room for ``rotating away'' the proton decay to the unobservable sector in the charged lepton case.

The particular robustness of the decay modes with neutrinos in the final states is subsequently reflected in the robustness of the total proton lifetime in $SO(10)$ unifications, see FIG.~\ref{figSO10-total}. This is due to the fact that, in this scenario, these partial widths are never significantly suppressed with respect to those into charged leptons.

\subsubsection{Flipped SU(5) unifications.}
As explained above, the lack of dedicated fits of the Yukawa structure for this class of unification models lead us to choosing a random starting point when the stability of this scenario with respect to the Planck-scale corrections was examined. Remarkably, the qualitative behaviour was independent of the starting point choice, hence, we believe that also the results regarding the flipped SU(5) unifications are worth presenting here.

\begin{figure}
\begin{center}
\includegraphics[width=\columnwidth]{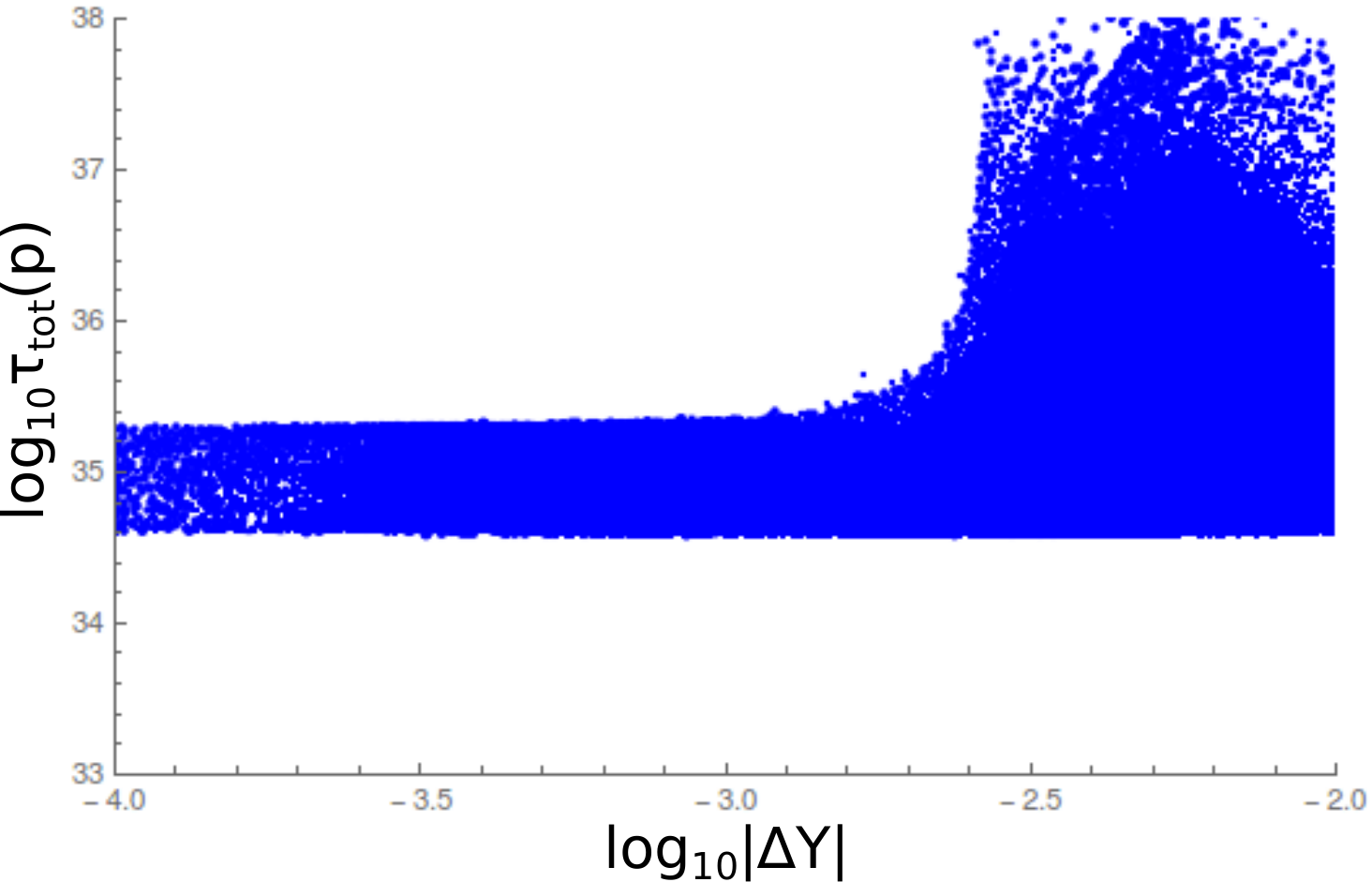}
\caption{Total proton lifetime in case of flipped SU(5) unifications where $D=D_C$ is assumed for the renormalizable-level ansatz for the Yukawa sector. The large spread of points with $|\Delta Y|$ only slightly below $10^{-2}$ would suggest that the proton decay can be indeed ``rotated away'' for this class of scenarios; however, if the minimal schemes are considered, the leading Planck-scale corrections are expected at $d=6$ level only, and for the corresponding $|\Delta Y|\sim 10^{-4}$ the total proton lifetime is constrained.}
\label{figFlipped}
\end{center}
\end{figure}

As shown in FIG.~\ref{figFlipped}, for $|\Delta Y|$ only slightly below $10^{-2}$ a significant instability of proton lifetime estimates based on the purely renormalizable structure was revealed. This supports the observation of \cite{Dorsner:2004jj} that the proton decay can be indeed rotated away in the flipped $SU(5)$ scenarios although at the renormalizable level the predictions seem to be rather robust (see the formulas \eqref{symFlipped}).

On the other hand, since in the simplest flipped $SU(5)$ models\footnote{Besides the original works \cite{GeorgiFlipped, BarrFlipped, DerFlipped} where the scalar sector is often not considered in detail, we have in mind, e.g., the fully realistic models including also non-zero neutrino masses like \cite{Das:2005eb} or \cite{Rodriguez:2013rma}} the unified gauge group is broken by a scalar representation charged with respect to the $U(1)_X$ gauge group, one easily finds that the $d=5$ Yukawa-affecting operators like ${\cal O}_{2}$ in (\ref{OpsList}) are absent. Consequently, the first Planck-induced structures that may generate uncontrolled shifts in the underlying Yukawa couplings emerge only at the $d=6$ level, hence, their size is expected to be of the order of $10^{-4}$. As can be seen in FIG.~\ref{figFlipped}, for such $|\Delta Y|$ the uncertainty in the total proton decay width becomes reasonably constrained, which means that, in the end, the proton lifetime estimates in the minimal flipped $SU(5)$ scenarios are particularly robust.

\subsubsection{SU(5) GUTs.}
Similarly as in the case of flipped $SU(5)$ unifications, also for ordinary $SU(5)$ we had to rely on a random renormalizable ansatz for the Yukawa matrices $Y_f'$ in our numerical analysis. Contrary to the flipped $SU(5)$ case, however, the realistic renormalizable models do not feature any symmetric Yukawa matrices (see Section~\ref{secYukRen}), hence, starting points with $U\neq U_C$ were assumed in general. For different initial settings of this type no common feature was observed -- even the total proton decay width could be spread over several orders of magnitude for certain choices of the starting point, although, on the other hand, for the settings with $U$ being close to $U_C$ the uncertainty coming from the unknown Planck-scale contributions was much smaller.\footnote{Let us mention as a curiosity that in case of the starting point featuring a symmetric up-type Yukawa matrix, i.e., with $U=U_C$,  the situation turns out to be even better than in the $SO(10)$ case since the Planck-scale induced uncertainty in the $p\to K^+\overline{\nu}$ partial width itself turns out to be constrained within less than one order of magnitude.}

\subsection{Remarks}
First, let us provide a hint on how the results presented in the plots above could be understood analytically. As stated in Lemma~\ref{lemmaEps} in Appendix~\ref{appMath}, if the first $n$ generalized eigenvalues of a matrix $Y_f'$ are of the order of $\mathcal{O}(\eps)$, then an $\mathcal{O}(\eps)$ correction to $Y_f'$ changes the $n\times n$ upper left blocks of the diagonalization matrices completely. This means that the uncertainty in the rotation matrices \eqref{rotace} and, hence, in the proton lifetime estimates qualitatively changes whenever the size of the perturbations crosses a generalized eigenvalue of $Y_f'$ (which is proportional to one of the fermion masses). Since the largest entries of the down-quark and charged-lepton Yukawa matrices corresponding to the $b$ and $\tau$ masses are around $5\times 10^{-3}$, for such values of $|\Delta Y|$ there appears the first ``step'' in the FIGs.~\ref{figSO10-nus},\ref{figSO10-total} (when perturbations with the size above this threshold are added to $Y_f'$'s the matrices $D,D_C,E,E_C$ can be changed completely; for $|\Delta Y|$ below this threshold only the upper left $2\times 2$ corner of these matrices may be significantly varied). The other ``step'' in plots of FIGs.~\ref{figSO10-nus},\ref{figSO10-total} corresponds to crossing the values of the Yukawa matrix entries corresponding to the $c$ and $\mu$ masses.

Finally, a comment is worth concerning the $SO(10)$ with the $10\oplus 126\oplus 54$ scalar sector. As mentioned in Section~\ref{secComentsMethod} this scenario is exceptional since both the renormalizable $Y_f$'s and the $d=5$ Planck-induced corrections $\Delta Y_f$'s are symmetric, hence, the flavour structure of the partial proton widths with neutrinos in the final state remains fully determined even if the non-renormalizable terms are included. 
However, even with this extra information at hand, the decay channels with the charged leptons in the final state still exhibit a numerical behaviour similar to the case with general $\Delta Y_f'$'s (see FIG.~\ref{figGold}), i.e., the spread in these partial widths remains rather large.

\section{Conclusions and outlook}\label{secConc}
In the current study, we have elaborated on the robustness of the gauge-boson-mediated contributions to the proton decay width in the simplest renormalizable unified models based on the $SO(10)$ and $SU(5)$ gauge groups with respect to several types of uncertainties, in particular those due to the presence of 
the Planck-scale induced operators altering the renormalizable-level Yukawa structure of specific models (such as ${\cal O}_2$ in the list (\ref{OpsList})). These perturbations, as small as they may seem in comparison to the typically huge effects inflicted by the notorious gauge-kinetic-form-changing $d=5$ operators (i.e., ${\cal O}_1$ in (\ref{OpsList})), are still significant enough to trigger large changes in the mixing matrices governing the relevant baryon and lepton number violating currents and, hence, cripple in principle the credibility of any of the existing renormalizable-level proton lifetime estimates.        

Remarkably enough, a thorough numerical analysis reveals vastly different levels of robustness of the relevant proton decay widths across different variants of the simplest $SO(10)$ and $SU(5)$ scenarios. Let us recapitulate the main observations that we managed to make here:

Unfortunately, for all scrutinized models, the individual decay channels with charged leptons in the final state were found to be prone to significant destabilisation  even for rather small Planck-induced perturbations.  
Typically, the inflicted theoretical uncertainties prevent, e.g., the ``golden'' channel $p\to \pi^0 e^+$ from discriminating efficiently among different scenarios (even with a fit to the renormalizable Yukawa structure at hand) unless the overall suppression associated to the relevant $d=5$ operators happens to be well under the $10^{-2}$ level  expected from the simple $M_G/M_{Pl}$ ratio (see FIG.~\ref{figGold}).

Concerning the $SO(10)$ scenarios, our numerical analysis reveals that, for all available renormalizable Yukawa fits within the minimal renormalizable models, the sum of the partial decay widths with neutrinos in the final state, and, consequently also the total proton lifetime turn out to be quite trustable from the flavour structure point of view even if the overall suppression factor associated to the Planck-scale  effects is as large as $10^{-2}$ (see FIGs.~\ref{figSO10-nus}~and~\ref{figSO10-total}). 
This applies, namely, to the minimal potentially realistic scenario with the GUT-scale symmetry breaking triggered by the $45$ scalar~\cite{Bertolini:2013vta,Kolesova:2014mfa} which, besides this feature, exhibits a spectacular level of robustness with respect to the gauge-kinetic effects associated to the ${\cal O}_1$ operator in the list (\ref{OpsList}).  
Hence, the leading Planck-scale induced theoretical uncertainties plaguing the renormalizable-level proton lifetime estimates in this scenario can be well within the ``improvement windows'' of the upcoming megaton-scale facilities such as Hyper-K or DUNE.
It is also worth mentioning that the alternative scenario in which $SO(10)$ is broken by 54 instead of 45 is yet more robust as far as the flavour strucure is concerned. On the other hand, it suffers from significant theoretical uncertainties in the overall BLNV scale determination which make it somewhat less attractive from the phenomenology point of view.   
  
Concerning the $SU(5)$-based scenarios, the flipped $SU(5)$ would be our primary choice as (in the minimal variants such as, e.g.,~\cite{Rodriguez:2013rma}) it typically exhibits a higher degree of robustness of the flavour structure governing the proton lifetime calculations due to the generic absence of the potentially dangerous Planck-induced corrections at the $d=5$ level (and it admits no ${\cal O}_1$-type operator either), see FIG.~\ref{figFlipped}. 
On the other hand,
we have nothing specific (and model independent) to say about the the robustness of the proton lifetime estimates made within the standard Georgi-Glashow scenario and/or its simple variants.

\subsection*{Acknowledgments}
The work of M.M. has been partially supported by the Marie-Curie Career Integration Grant within the 7th European Community Framework Programme
FP7-PEOPLE-2011-CIG, contract number PCIG10-GA-2011-303565, by the Research proposal MSM0021620859 of the Ministry of Education, Youth and Sports of the Czech Republic , by the Foundation for support of science and research ``Neuron'' and by the Grant agency of the Czech Republic, project no. 17-04902S. The work of H.K. was supported by the Grant Agency of the Czech Technical University in Prague, grant No. SGS13/217/OHK4/3T/14. We are grateful to Renato Fonseca for his insightful comments on the flavour-space symmetry properties of $d>4$ Yukawa-type operators.

\appendix
\section{Prefactors entering the partial proton decay widths}\label{appWidths}
Let us complete the formulas for the partial proton decay widths \eqref{gammapi0}-\eqref{gammaK+} by the definition of the flavour independent prefactors. For the sake of continuity with the previous works we use the parametrization used in \cite{proton} based on the chiral Lagrangian:
\begin{align}
C_\pi&=\frac{m_p}{16\pi f_\pi^2}A_L^2|\alpha|^2(1+D+F)^2\nn\\	
C_\eta&=\frac{(m_p^2-m_\eta^2)^2}{48\pi m_p^3 f_\pi^2}A_L^2|\alpha|^2(1+D-3F)^2\nn\\
C_K&=\frac{(m_p^2-m_K^2)^2}{8\pi m_p^3 f_\pi^2}A_L^2|\alpha|^2\,,\;\; B_1=1+\frac{m_p}{m_B}(D-F)\nn\\
B_2&=\frac{2m_p}{3m_B}D\qquad B_3=1+\frac{m_p}{3m_B}(D+3F)\nn
\end{align}
where $m_p$, $m_\eta$ and $m_K$ denote the proton, $\eta$ and kaon mass, respectively, $m_B$ is an average baryon mass ($m_B\approx m_\Sigma \approx m_\Lambda$),  $f_\pi$ is the pion decay constant, $|\alpha|$, $D$ and $F$ are the parameters of the chiral Lagrangian, and $A_L$ takes into account the renormalization from $M_Z$ to 1~GeV.

On the other hand, the recent lattice computations \cite{Aoki:2017puj} predict directly the individual matrix elements without the use of the chiral Lagrangian. These results can be, however, translated to the chiral Lagrangian parametrization (see, e.g., Appendix A of \cite{Aoki:2017puj}) and the value of the parameter $\alpha$ is then inferred.\footnote{Let us note that only the matrix elements of the RL type like $\langle\pi^0|(ud)_R u_L|p\rangle$ are relevant for the vector-boson-mediated proton decay, hence, only the parameter $\alpha$ enters the formulas \eqref{gammapi0}-\eqref{gammaK+}.} Let us stress that the possible change in the value of this multiplicative factor given by the improvement of the lattice computations does not affect our results qualitatively, the points in all the plots would be merely shifted in a uniform way.

\section{Matrix diagonalization}\label{appMath}
Let a complex matrix $Y$ be diagonalized by a biunitary transformation
\begin{equation}\label{defU}
U_C^T Y U = Y_d
\end{equation}
where $Y_d$ is a real non-negative diagonal matrix consisting of the so-called generalized eigenvalues of $Y$. Since the main issue of this paper is the sensitivity of the unitary matrices $U,U_C$ to the small perturbations in the matrix $Y$, let us mention here few mathematical results on this problem.

As a first step, however, let us remind the reader about the way in which the matrices $U$ and $U_C$  in \eqref{defU} are constructed. Since $Y^\dagger Y$ is a hermitian matrix, it can be diagonalized as 
\begin{equation}\label{defD}
U^\dagger Y^\dagger Y U = D
\end{equation}
where $D$ is a real non-negative diagonal matrix. The diagonal matrix in \eqref{defU} is then defined as $Y_d = \sqrt{D}$ and if its entries are non-zero, then the unitary matrix $U_C$ can be defined as
\begin{equation}\label{defUC}
U_C^\ast = Y U Y_d^{-1}.
\end{equation}
Let us note, however, that the matrices $U$ and $U_C$ are not defined uniquely and the level of this ambiguity depends on the shape of $Y_d$: \\
$i)$ For non-degenerate and non-zero diagonal entries of $Y_d$, the ambiguity in $U$ in \eqref{defD} amounts to $U\to U P$ with $P$ being a diagonal unitary matrix. The matrix $U_C$ in \eqref{defUC} is then accordingly transformed as $U_C\to U_C P^\ast$. \\
$ii)$ If w.l.o.g. $Y_d^{11}=Y_d^{22}=\dots=Y_d^{nn}\neq 0$, then $U\to U U_n,\, U_C\to U_C U_n^{\ast}$
is allowed where the upper left corner of $U_n$ is formed by an $n\times n$ unitary block, $U_n^{jj}=e^{i\phi_j}$ for $j>n$ and $U^n_{ij}=0$ otherwise. \\
$iii)$ If, finally, w.l.o.g. $Y_d^{11}=Y_d^{22}=\dots=Y_d^{nn}= 0$, then the definition \eqref{defUC} of $U_C$ can not be applied and the relation 
\begin{equation}\label{defDUC}
U_C^T Y Y^\dagger U_C^* = D
\end{equation}
has to be used instead. The ambiguity in the definition of the rotation matrices than reads
\begin{equation}\label{ambU}
U\to U U_n,\quad U_C\to U_C U_{Cn}^{\ast}
\end{equation}
with the same structure of $U_n, U_{Cn}$ as in point $ii)$. Here, however, the upper left $n\times n$ blocks of $U_n$, $U_{Cn}$ are uncorrelated and the phases of the diagonal entries $U_n^{jj}, U_{Cn}^{jj}$ for $j>n$ have to be adjusted in such a way that $Y_d$ in \eqref{defU} is real and non-negative.

 When $Y$ is perturbed by an $\mathcal{O(\eps)}$ amount with $\eps$ being a small parameter, one would naively expect also $\mathcal{O(\eps)}$ changes in the unitary matrices $U, U_C$ in \eqref{defU}. The following statement confirms this expectation under certain assumptions.
 
\begin{lemma}\label{lemma1}
Let $Y$ be a complex matrix diagonalized by the biunitary transformation \eqref{defU}
with $Y_d^{ii}\sim \O(1)$ $\forall i$ and $Y_{d}^{ii}\neq Y_{d}^{jj}$ $\forall i\neq j$. Further, let $X$ be an arbitrary complex matrix and let us define
$\tilde{X}=U_C^T X U$. Then
\begin{equation}\label{pertY}
\tilde{U}_C^T(Y+\eps X)\tilde{U} = Y_d + \eps R_d + \O(\eps^2)
\end{equation}
where $R_d$ is a real diagonal matrix with $R_d^{ii}=\mathrm{Re} \tilde{X}^{ii}$ and 
\begin{equation}\label{pertU}
\tilde{U}=U (\unit+\eps Z) + \O(\eps^2), \quad \tilde{U}_C=U_C (\unit+\eps Z_C)+\O(\eps^2)
\end{equation}
for some antihermitian matrices $Z$ and $Z_C$.
\end{lemma} 
\emph{Proof.} Let us assume that the generalized eigenvalues of $Y+\eps X$ and also the corresponding rotation matrices $U, U_C$  are changed by the $\mathcal{O}(\eps)$ values, i.e., the diagonalization of $Y+\eps X$ follows \eqref{pertY} with $\tilde{U}, \tilde{U}_C$ of the form \eqref{pertU}. For $\tilde{U}, \tilde{U}_C$ to be unitary (up to $\O(\eps^2)$ terms), $Z^\dagger=-Z$, $Z_C^\dagger=-Z_C$ has to be satisfied, hence, $Z, Z_C$ are indeed antihermitian. In order to prove the lemma, the matrices $R_d$ and $Z, Z_C$ will be explicitly constructed.

According to \eqref{defD}, the matrix $\tilde{U}$ is defined by
$$
\left(Y+\varepsilon X\right)^\dagger \left(Y+\varepsilon X\right)=\tilde{U} \left(Y_d + \eps R_d+\mathcal{O}(\eps^2)\right)^2 \tilde{U}^\dagger.
$$
If the shape of $\tilde{U}$ \eqref{pertU} is plugged in, then the equality of the $\mathcal{O}(\eps)$ terms yields
$$Y^\dagger X+X^\dagger Y = U\left( Z Y_d^2  - Y_d^2 Z + 2 Y_d R_d\right) U^\dagger.$$
Multiplying this relation by $U^\dagger$ from the left and by $U$ from the right one obtains
\begin{equation}\label{pr2}
Y_d \tilde{X} + \tilde{X}^\dagger Y_d = Z Y_d^2 - Y_d^2 Z + 2 Y_d R_d
\end{equation}
where $\tilde{X}=U_C^T X U$ was defined. Taking the diagonal elements of this matrix relation, one obtains (after dividing by $Y_d^{jj}$)
$\tilde{X}^{jj} + \tilde{X}^{jj\ast} = 2 R_d^{jj}$
which shows that indeed $R_d^{jj} = \mathrm{Re} \tilde{X}^{jj}$ as stated in the lemma. On the other hand, the off-diagonal elements of the matrix relation \eqref{pr2} allow to compute the matrix $Z$:
\begin{equation}\label{defZ}
Z^{ij} =[{Y_d^{ii} \tilde{X}^{ij}+\tilde{X}^{ji\ast} Y_d^{jj}}]/[{(Y_d^{jj})^2 - (Y_d^{ii})^2}]\qquad i\neq j.
\end{equation} 
Finally, any purely imaginary number can be chosen as the diagonal entries of $Z$ which corresponds to the ambiguity in the definition of the rotation matrices mentioned in point $i)$ above. 

The matrix $Z_C$ can be constructed analogously when \eqref{defDUC} is taken into account and
\begin{equation}\label{defZC}
Z_C^{ij} = [{Y_d^{ii} \tilde{X}^{ji}+\tilde{X}^{ij\ast} Y_d^{jj}}]/[
{(Y_d^{jj})^2 - (Y_d^{ii})^2}]\qquad i\neq j
\end{equation}
is obtained for non-diagonal elements.

Finally, one can plug in the shape of $\tilde{U}$ and $\tilde{U}_C$ into the formula \eqref{pertY} and the $\mathcal{O}(\eps)$ part of this relation then yields
\begin{equation}\label{pr3}
Z_C^T Y_d + Y_d Z +\tilde{X} = R_d.
\end{equation}
If the formulas for $Z$ and $Z_C$ are plugged in, one can easily check that indeed the off-diagonal elements of the left-hand side are equal to zero. Moreover, if the imaginary part of the diagonal elements is evaluated, one obtains
$$\left(Z^{jj} + Z_C^{jj}\right)Y_d^{jj} + i\, \mathrm{Im}\tilde{X}^{jj}=0$$
which fixes the (purely imaginary) diagonal entries of $Z_C$.
\endproof

It is now easy to understand why the above described construction breaks down when $Y_d^{11}=Y_d^{22}=\dots=Y_d^{nn}=0$. No information about $Z^{ij}$ for $i,j\leq n$ can be obtained from \eqref{pr2} (and analogously, no information about $Z_C^{ij}$ for $i,j\leq n$ is available). Moreover, \eqref{pr3} simplifies to $\tilde{X}^{ij} = R_d^{ij}$ for $i,j\leq n$, hence, the upper left $n\times n$ block of $\tilde{X}$ has to be diagonal. This can be ensured thanks to the ambiguity \eqref{ambU} in the definition of the $U$ and $U_C$ matrices in \eqref{defU}, with $U_n$ and $U_{Cn}$ being chosen in such a way that
\begin{equation}\label{liftAmb}
\left(U_{Cn}^T \tilde{X} U_n\right)^{ij} =\left(U_{Cn}^T U_C^T X U U_n\right)^{ij} = 0 \quad i\neq j,\;i,j\leq n.
\end{equation}
The shape of the perturbed rotation matrices \eqref{pertU} given in Lemma~\ref{lemma1} can be, hence, still used, however, the ambiguity \eqref{ambU} is lifted.

Furthermore, let us consider the setting with $\mathcal{O}(\eps)$ generalized eigenvalues, more precisely let
\begin{equation}\label{defUeps}
U_C^T Y U = Y_d + \eps\Lambda_n
\end{equation}
where $Y_d$ and $\Lambda_n$ are diagonal matrices with $Y_d^{jj}=0$ for $j\leq n$ and $\Lambda_n^{jj}=0$ for $j> n$. In order to illustrate the effect of an $\mathcal{O}(\eps)$ perturbation in this case, let us define
$$Y_0= Y-\eps U_C^\ast \Lambda_n U^\dagger$$
which obviously has first $n$ generalized eigenvalues equal to zero and the considerations of the previous paragraph can be applied. $Y$ can be then viewed as a perturbation of $Y_0$ by an $\mathcal{O}(\eps)$ term which lifts the ambiguity in the definition of $U$ and $U_C$ as described in \eqref{liftAmb}. Similarly, $Y+\eps X$ can be understood as a different perturbation of $Y_0$ and clearly, in general, $U$ and $U_C$ are fixed in a different way. It is then easy to check the following statement.

\begin{lemma}\label{lemmaEps}
Let $Y$ be a complex matrix diagonalized by a biunitary transformation as in \eqref{defUeps}, let $X$ be an arbitrary complex matrix and $\eps$ a small parameter. Then
\begin{equation}\label{pertY2}
\tilde{U}_C^T(Y+\eps X)\tilde{U} = Y_d + \eps \tilde{R}_d+ \O(\eps^2)
\end{equation}
where $\tilde{R}_d$ is a diagonal matrix and the rotation matrices may be written in the form
\begin{equation}
\tilde{U}=U U_n (\unit+\eps Z)+ \O(\eps^2) \quad \tilde{U}_C=U_C U_{Cn} (\unit+\eps Z_C)+ \O(\eps^2)
\end{equation} for some antihermitian matrices $Z$, $Z_C$ and unitary matrices $U_n$, $U_{Cn}$ where $U_n^{ij}=U_{Cn}^{ij}=0$ for $i,j>n$, $i\neq j$.
\end{lemma}

In our work we are interested in perturbations preserving the generalized eigenvalues of the original matrix. The above results on the shape of the diagonalization matrices can be used also in this case due to the following simple observation.

\begin{lemma}\label{pokus}
For any complex matrix $X$ there exists a complex matrix $X'$ such that the generalized eigenvalues of $Y+\eps X'$ differ from the generalized eigenvalues of $Y$ by at most $\mathcal{O}(\eps^2)$ terms and, at the same time, $Y+\eps X$ and $Y+\eps X'$ are diagonalized by the same biunitary transformation.
\end{lemma}
\emph{Proof.} Looking at the relation \eqref{pertY2} it is enough to define $X'=X - \tilde{U}_C^\ast \tilde{R}_d \tilde{U}^\dagger.$
\endproof

\bibliographystyle{utphysmod}
\bibliography{GUT}

\end{document}